\documentclass[twocolumn,showpacs]{revtex4}
\usepackage{epsfig}
\usepackage{amsmath}

\begin{document}

\author{F. Sagu\'es}
\affiliation{Departament de
Qu\'imica F\'isica, Universitat de Barcelona, Mart\'i i Franqu\`es
1, E-08028, Barcelona, Spain}
\author{V.P. Shkilev}
\affiliation{Institite of Surface Chemistry, National Academy of Sciences of Ukraine, 
UA-03164 Kiev, Ukraine}
\author{I.M. Sokolov}
\affiliation{Institut f\"ur Physik,
Humboldt-Universit\"at zu Berlin, Newtonstra\ss e 14, 12489 Berlin, Germany;
e-mail: igor.sokolov@physik.hu-berlin.de}

\title{Reaction-Subdiffusion Equations for the A $\rightleftharpoons$ B Reaction}

\begin{abstract}
We consider a simple linear reversible isomerization reaction
$\mathrm{A} \rightleftharpoons \mathrm{B}$ under subdiffusion
described by continuous time random walks (CTRW). The reactants'
transformations take place independently on the motion and are
described by constant rates. We show that the form of the ensuing
system of mesoscopic reaction-subdiffusion is somewhat unusual: the
equation giving the time derivative of one reactant concentration, say
$A(x,t)$, contains the terms depending not only on $\Delta A$, but
also on $\Delta B$, i.e. depends also on the transport operator of
another reactant. Physically this is due to the fact that several
transitions from A to B and back may take place at one site before the
particle jumps.
\end{abstract}

\maketitle

There are several reasons to discuss in detail the structure of
mesoscopic kinetic equations describing the behavior of a simple reversible
isomerization reaction $\mathrm{A} \rightleftharpoons \mathrm{B}$
under subdiffusion. 

Many phenomena in systems out of equilibrium can be described within a
framework of reaction-diffusion equations. Examples can be found in
various disciplines ranging from chemistry and physics to
biology. Both reaction-diffusion systems with normal and anomalous
diffusion have been extensively studied over the past decades.
However, for the latter, a general theoretical framework which would
hold for all kinds of reactions is still absent. The reasons for
subdiffusion and therefore its properties can be different in systems
of different kind; we concentrate here on the situations when such
subdiffusion can be adequately described by continuous-time random
walks (CTRW). In CTRW the overall particle's motion can be considered
as a sequence of jumps interrupted by waiting times, the case
pertinent to many systems where the transport is slowed down by
obstacles and binding sites.  In the case of anomalous diffusion these
times are distributed according to a power law lacking the mean. The
case of exponential distribution, on the other hand, corresponds to a
normal diffusion.  On the microscopic level of particles' encounter
the consideration of subdiffusion does not seem to be problematic,
although it has posed several interesting questions
\cite{micro1,micro2,micro3,micro4,micro5}. However, these microscopic
approaches cannot be immediately adopted for description of spatially
inhomogeneous systems, which, in the case of normal diffusion, are
successfully described within the framework of reaction-diffusion
equations. To discuss such behavior under subdiffusion many authors
used the kind of description where the customary reaction term was
added to a subdiffusion equation for concentrations to describe such
phenomena as a reaction front propagation or Turing instability
\cite{M1,M2,HW1,HW2,HLW,LHW}.
 
The results of these works were jeopardized after it was shown in
Ref.\cite{SSS1} that these procedure does not lead to a correct
description even of a simple irreversible isomerization reaction
$\mathrm{A} \rightarrow \mathrm{B}$. The transport operator describing
the subdiffusion is explicitly dependent on the properties of
reaction, which stems from an essentially very simple observation that
only those particles jump (as A) which survive (as A).

The properties of the reaction depend strongly on whether the reaction
takes place only with the step of the particle, or independently on
the particles' steps, and moreover, whether the newborn particle
retains the rest of it previous waiting time or is assigned a new one
\cite{HLW1,NeNe}.  Here we consider in detail the following situation:
The $\mathrm{A} \rightleftharpoons \mathrm{B}$ transformations take
place independently on the particles' jumps; the waiting time of a
particle on a site is not changed by the reaction, both for the the
forward and for the backward transformation.  As a motivation for such
a scheme we can consider the reaction as taking place in an aqueous
solution which soaks a porous medium (say a sponge or some geophysical
formation).  If sojourn times in each pore are distributed
according to the power law, the diffusion on the larger scales is anomalous; 
on the other hand, the reaction within each
pore follows usual kinetics. We start by putting a droplet containing,
say, only A particles somewhere within the system and follow the
spread and reaction by measuring the local A and B concentrations.

Stoichiometry of the chemical reaction implies the existence of a
conservation law.  In the case of the $\mathrm{A} \rightleftharpoons
\mathrm{B}$ it is evident that the overall number of particles is
conserved. If the isomerization takes place independently on the
particle's motion, then the evolution of the overall concentration
$C(x,t)=A(x,t)+B(x,t)$, where $A(x,t)$ and $B(x,t)$ are the local
concentrations of A and B particles respectively, is not influenced
by the reaction, and has to follow the simple subdiffusion equation
\[
\dot{C}(x,t) = K_\alpha \;_0D_t^{1-\alpha} \Delta C
\]
as it should be. On the other hand, neither the result of the treatment in
Ref.\cite{SSS2} nor the result of Ref.\cite{YH} reproduce this
behavior which is a consequence of the fundamental stoichiometry. In the
work \cite{SSS2} (where two of the authors of the present report were
involved) it was implicitly assumed that the back reaction can
only take place on a step of a particle, without discussing this
assumption. The corresponding description lead to the expressions
which could not be cast in a form resembling the reaction-diffusion
equations at all. The more general approach of Ref.\cite{YH},
definitely correct for irreversible reactions, also fails to reproduce
this local conservation law and thus is inappropriate for the
description of reversible reactions under the conditions discussed.  
According to Ref.\cite{NeNe} the approach of Ref.\cite{YH} implies that
the waiting time after each reaction is assigned anew, which makes a
large difference in the reversible case.

As a step on the way to understanding the possible form of the
reaction-subdiffusion equations we consider in what follows the
simplest linear reversible scheme where each step can be explicitly
checked.  We show that the form of the corresponding equations is
somewhat unusual, which emphasizes the role of coupling between the
reaction and transport in reaction-subdiffusion kinetics. Actually,
the equation giving the time derivative of one reactant concentration,
say $A(x,t)$, contains the terms depending not only on $\Delta A$, but
also on $\Delta B$, i.e. depends also on the transport operator of
another reactant. Physically this is due to the fact that several
transitions from A to B and back may take place at one site before the
particle jumps. This dependence disappears only in the Markovian case
due to vanishing of the corresponding prefactor.

Following the approach of  Ref. \cite{SSS1,SSS2} we describe the
behavior of concentrations in the discrete scheme by the
following equations:
\begin{eqnarray*}
&& \dot{A_i}(t)=-I_i(t)+\frac{1}{2} I_{i-1}(t)+\frac{1}{2} I_{i+1}(t)-k_1 A(t)+k_2 B(t) \\
&& \dot{B_i}(t)=-J_i(t)+\frac{1}{2} J_{i-1}(t)+\frac{1}{2} J_{i+1}(t)+k_1 A(t)-k_2 B(t) 
\end{eqnarray*}
where $I(t)$ is the loss flux of A-particles on site $i$ and $J(t)$ is
the corresponding loss flux for B-particles at site $i$. In the
continuous limit the equations read as
\begin{eqnarray}
\dot{A}(x,t)&=&\frac{a^2}{2}\Delta I(x,t)-k_1 A(x,t)+k_2 B(x,t) \\
\dot{B}(x,t)&=&\frac{a^2}{2}\Delta J(x,t)+k_1 A(x,t)-k_2 B(x,t).
\label{Cont} 
\end{eqnarray}
We now use the conservation laws for A and B particles to obtain the
equations for the corresponding fluxes. The equations for the
particles' fluxes on a given site in time domain (the index $i$ or the coordinate
$x$ is omitted) are:
\begin{widetext}

\begin{eqnarray}
I(t) &=& \psi(t) P_{AA}(t) A(0)
+ \int_0^t \psi(t-t') P_{AA}(t-t')
\left[ I(t') + k_1 A(t') - k_2B(t') +\dot{A}(t') \right] \nonumber \\
&+& \psi(t) P_{BA}(t) B(0) 
+ \int_0^t \psi(t-t') P_{BA}(t-t')
\left[ J(t') - k_1 A(t') + k_2B(t') +\dot{B}(t') \right]
\label{EqA}
\end{eqnarray}
for A-particles, and
\begin{eqnarray*}
J(t) &=& \psi(t) P_{BB}(t) B(0)
+ \int_0^t \psi(t-t') P_{BB}(t-t')
\left[ J(t') - k_1 A(t') + k_2B(t') +\dot{B}(t') \right] \\
&+& \psi(t) P_{AB}(t) A(0) 
+ \int_0^t \psi(t-t') P_{AB}(t-t')
\left[ J(t') + k_1 A(t') - k_2B(t') +\dot{A}(t') \right]
\end{eqnarray*}
for B-particles. 
\end{widetext}

The explanation of the form of e.g. Eq.(\ref{EqA}) is as follows:
An A-particle which jumps from a give site at time $t$ either was there
as A from the very beginning, and jumps as A probably having changed its
nature several time in between, or came later as A and jumps as A, or 
was there from the very beginning as B and leave the site as A, etc. 
 Here $P_{AA}$, $P_{AB}$, $P_{BA}$ and $P_{BB}$ 
are the survival/transformation probabilities, i.e. the probability that a particle coming
to a site as A at $t=0$ leaves it at time $t$ as A (probably having changed its nature
from A to B and back in between), the probability that a particle coming
to a site as A at $t=0$ leaves it at time $t$ as B, the probability that a particle coming
to a site as B at $t=0$ leaves it at time $t$ as A, and the probability that a particle coming
to a site as B at $t=0$ leaves it at time $t$ as B: 
\begin{eqnarray}
P_{AA}(t)&=&\frac{k_2}{k_1+k_2}+\frac{k_1}{k_1+k_2}e^{-(k_1+k_2)t} \nonumber \\
P_{BA}(t)&=&\frac{k_2}{k_1+k_2}-\frac{k_2}{k_1+k_2}e^{-(k_1+k_2)t} \nonumber \\
P_{BB}(t)&=&\frac{k_1}{k_1+k_2}+\frac{k_2}{k_1+k_2}e^{-(k_1+k_2)t} \nonumber \\
P_{AB}(t)&=&\frac{k_1}{k_1+k_2}-\frac{k_1}{k_1+k_2}e^{-(k_1+k_2)t}
\end{eqnarray}
These are given by the solution of the classical reaction kinetic equations
\begin{eqnarray}
\dot{A}(t)&=&-k_1A(t)+k_2B(t) \nonumber \\
\dot{B}(t)&=&k_1A(t)-k_2B(t).
\end{eqnarray}
The values of $P_{AA}$, $P_{AB}$ are given by the solutions $P_{AA}(t)=A(t)$ and $P_{AB}(t)=B(t)$ 
under initial conditions $A(0)=1,\; B(0)=0$, and the values of $P_{BA}$ and $P_{BB}$
are given by  $P_{BA}(t)=A(t)$ and $P_{BB}(t)=B(t)$ under initial conditions $A(0)=0,\; B(0)=1$.  

In the Laplace domain we get:
\begin{eqnarray}
I(u)&=&\psi_1(u)\left[ I(u)+k_1A(u)-k_2B(u)+uA(u)\right] \nonumber \\
&& +\psi_2(u)\left[J(u)-k_1A(u)+k_2B(u)+uB(u)\right] \nonumber \\
J(u)&=&\psi_3(u)\left[J(u)-k_1A(u)+k_2B(u)+uB(u)\right] \label{system1}\\
&& +\psi_4(u)\left[ I(u)+k_1A(u)-k_2B(u)+uA(u)\right] \nonumber
\end{eqnarray}
where $\psi_1(u),\psi_2(u), \psi_3(u)$ and $\psi_4(u)$ are the Laplace transforms of 
$\psi_1(t)=\psi(t)P_{AA}(t)$, $\psi_2(t)=\psi(t)P_{BA}(t)$, $\psi_3(t)=\psi(t)P_{BB}(t)$ and $\psi_4(t)=\psi(t)P_{AB}(t)$, 
respectively.

Using shift theorem we can get the representations of $\psi_i$ in the Laplace domain. They
read:
\begin{eqnarray}
\psi_1(u)&=&\frac{k_2}{k_1+k_2}\psi(u)+\frac{k_1}{k_1+k_2}\psi(u+k_1+k_2)\nonumber \\
\psi_2(u)&=&\frac{k_2}{k_1+k_2}\psi(u)-\frac{k_2}{k_1+k_2}\psi(u+k_1+k_2)\nonumber \\
\psi_3(u)&=&\frac{k_1}{k_1+k_2}\psi(u)+\frac{k_2}{k_1+k_2}\psi(u+k_1+k_2)\nonumber \\
\psi_4(u)&=&\frac{k_1}{k_1+k_2}\psi(u)-\frac{k_1}{k_1+k_2}\psi(u+k_1+k_2)
\end{eqnarray}
The system of linear equations for the currents, Eqs.(\ref{system1}), then has the
solution 
\begin{eqnarray*}
I(u)=a_{11}(u) A(u)+ a_{12}(u) B(u) \\
J(u)=a_{21}(u) A(u)+ a_{22}(u) B(u)
\end{eqnarray*}
with the following values for the coefficients:
\begin{eqnarray*}
a_{11} &=& \frac{1}{k_1+k_2}\frac{1}{1+\phi\psi-\psi-\phi} \\
 &\times& \left[-\phi\psi\left(k_1k_2+u(k_1+k_2)+k_1^2\right) \right. \\
&+& \left. \phi k_1(u+k_1+k_2)+\psi k_2 u\right], 
\end{eqnarray*}

\begin{eqnarray*}
a_{21} &=& \frac{k_1}{k_1+k_2}\frac{1}{1+\phi\psi-\psi-\phi} \\
&\times& \left[\phi\psi (k_1+k_2 )+(\psi-\phi)u-(k_1+k_2)\phi\right],
\end{eqnarray*}
and with the two other coefficients, $a_{12}$ and $a_{22}$ differing from
$a_{21}$ and $a_{11}$ by interchanging $k_1$ and $k_2$.  Here $\psi \equiv \psi(u)$ and 
$\phi \equiv \psi(u+k_1+k_2)$. 

For the exponential
waiting time density $\psi(t)=\tau^{-1}\exp(-t/\tau)$ the
corresponding values get 
\begin{eqnarray*}
a_{11}&=&a_{22}=1/\tau \\
a_{12}&=&a_{21}=0,
\end{eqnarray*}
and the system of equations for the concentrations in the continuous
limit, Eqs.(\ref{Cont}), reduces to the customary system of reaction-diffusion
equations. For the case of the power-law distributions $\psi(t) \simeq
t^{-1-\alpha}$ the Laplace transform of the waiting time PDF is
$\psi(u) \simeq 1-cu^\alpha$ for small $u$, with $c=\tau^\alpha \Gamma(1-\alpha)$, so that
\begin{eqnarray*}
a_{11}&=&\frac{c^{-1}}{k_1+k_2} \left[ k_2u^{1-\alpha}+k_1(u+k_1+k_2)^{1-\alpha} \right.\\
&-& \left. ck_1(k_1+k_2) -cu(k_1+k_2)\right] \\
a_{22}&=&\frac{c^{-1}}{k_1+k_2} \left[ k_1u^{1-\alpha}+k_2(u+k_1+k_2)^{1-\alpha} \right. \\
 &-& \left. ck_2(k_1+k_2) -cu(k_1+k_2)\right] \\
a_{21}&=&\frac{c^{-1}}{k_1+k_2} \left[ k_1u^{1-\alpha}-k_1(u+k_1+k_2)^{1-\alpha} \right. \\
 &+& \left. ck_1(k_1+k_2) \right] \\
a_{12}&=&\frac{c^{-1}}{k_1+k_2}\left[ k_2u^{1-\alpha}-k_2(u+k_1+k_2)^{1-\alpha} \right. \\
 &+& \left. ck_2(k_1+k_2) \right].
\end{eqnarray*}
Now we turn to the case of long times and relatively slow reactions,
so that all parameters, $u$, $k_1$ and $k_2$ can be considered as
small. In this case, for $\alpha<1$, the leading terms in all these
parameters are the first two terms in each of the four equations, and
the other terms can be neglected. In the time domain the operator
corresponding to $u^{1-\alpha}$ is one of the fractional derivative
$_0D_t^{1-\alpha}$, and the operator corresponding to
$(u+k_1+k_2)^{1-\alpha}$ is the transport operator of Ref.\cite{SSS2},
$_0\mathcal{T}_t^{1-\alpha}(k_1+k_2)$ with
$_0\mathcal{T}_t^{1-\alpha}(k) = e^{-kt} \;_0D_t^{1-\alpha} e^{kt}$.
Introducing the corresponding equations for the currents into the
balance equations for the particle concentrations we get:
\begin{widetext}
\begin{eqnarray}
\dot{A}(x,t)&=&K_\alpha \left[\frac{k_2}{k_1+k_2} \;_0D_t^{1-\alpha}+
\frac{k_1}{k_1+k_2} \;_0\mathcal{T}_t^{1-\alpha}(k_1+k_2) \right] \Delta A(x,t)\nonumber \\
&+&K_\alpha \left[\frac{k_2}{k_1+k_2} \;_0D_t^{1-\alpha}-
\frac{k_2}{k_1+k_2} \;_0\mathcal{T}_t^{1-\alpha}(k_1+k_2) \right] \Delta B(x,t) 
-k_1 A(x,t)+k_2 B(x,t) \label{ConcA} \\
\dot{B}(x,t)&=& K_\alpha \left[\frac{k_1}{k_1+k_2} \;_0D_t^{1-\alpha}-
\frac{k_1}{k_1+k_2} \;_0\mathcal{T}_t^{1-\alpha}(k_1+k_2) \right] \Delta A(x,t) \nonumber \\
&+&K_\alpha \left[\frac{k_1}{k_1+k_2} \;_0D_t^{1-\alpha}+
\frac{k_2}{k_1+k_2} \;_0\mathcal{T}_t^{1-\alpha}(k_1+k_2) \right] \Delta B(x,t) 
+k_1 A(x,t)-k_2 B(x,t). \label{ConcB}
\end{eqnarray}
\end{widetext}
Note also that the equation for $C(x,t) = A(x,t)+B(x,t)$ following from summing up the 
Eqs.(\ref{ConcA}) and (\ref{ConcB}) is a simple subdiffusion equation 
\[
\dot{C}(x,t) = K_\alpha \;_0D_t^{1-\alpha} \Delta C
\]
as it should be. On the other hand, neither the result of the
treatment in Ref.\cite{SSS2} nor the result of Ref.\cite{YH} reproduce
this behavior which is a consequence of the fundamental conservation
law prescribed by the stoichiometry of reaction.

Note that this system still holds for $\alpha=1$ when both the
fractional derivative $\;_0D_t^{1-\alpha}$ and the transport operator
$\;_0\mathcal{T}_t^{1-\alpha}(k_1+k_2)$ are unit operators. In this
case the usual system of reaction-diffusion equations is restored:
\begin{eqnarray*}
\dot{A}(x,t)&=& K \Delta A(x,t)-k_1 A(x,t)+k_2 B(x,t) \\
\dot{B}(x,t)&=& K \Delta B(x,t)+k_1 A(x,t)-k_2 B(x,t). 
\end{eqnarray*}

Let us summarize our findings. We considered the system of mesoscopic
(reaction-subdiffusion) equations describing the kinetics of a
reversible isomerization $\mathrm{A} \rightleftharpoons \mathrm{B}$
taking place in a subdiffusive medium. When the waiting times of the
particles are not assigned anew after their transformations (i.e. when
the overall concentration of reactants is governed by the simple
subdiffusion equation), this reaction is described by a rather unusual
system of reaction-subdiffusion equations having a form which was up
to our knowledge not discussed before: Each of the equations, giving
the temporal changes of the corresponding concentrations, depends on
the Laplacians of \textit{both} concentrations, $A$ and $B$ (not only
on the same one, as in the case of normal diffusion). This is a rather
unexpected situation especially taking into account the fact that our
reaction is practically decoupled from the transport of particles. The
form reduces to a usual reaction-diffusion form for normal diffusion
(due to cancellations). It is important to note that the physical
reason of the appearance of such a form is the possibility of several
transformations $\mathrm{A} \to \mathrm{B} \to \mathrm{A} \to
\mathrm{B} \, ...$ during one waiting period, and that such
possibilities have to be taken into account also for more complex
reactions including reversible stages.

\textbf{Acknowledgement:} IMS gratefully acknowledges financial
support by DFG within SFB 555 joint research program. FS
acknowledges financial support from MEC under project FIS 2006 -
03525 and from DURSI under project 2005 SGR 00507.

\end{document}